# Position Sensing Errors in Synchronous Motor Drives

**Prerit Pramod**, *Senior Member*, IEEE
Control Systems Engineering, MicroVision, Inc.
*Email*: preritpramod89@gmail.com; preritp@umich.edu; prerit_pramod@microvsion.com

**Abstract:** Non-ideal position estimation results in degraded performance of synchronous motor drive systems due to reduction of the average capability of the drive as well as torque harmonics of different orders. The signature and extent of the performance degradation is further dependent, quite significantly, on the current control architecture, i.e., feedforward or feedback control, employed. This paper presents a comprehensive analysis of non-idealities or errors in position estimation and their effects on the control performance of synchronous motor drives. Analytical models capturing the error in various signals caused by position sensing errors in the drive system for different control architectures are presented and are validated with simulation and experimental results on a prototype permanent magnet synchronous motor drive.

## Introduction

Modern motion control systems are required to meet stringent performance, reliability and safety standards, particularly those employed in the automotive [1]–[12] and medical [13] industries. Motion control is achieved in part by utilizing electric motor drive systems to generate the requisite electromagnetic torque or force [14]–[17] to drive mechanical actuators. A wide variety of electric machine topologies ranging from synchronous [18]–[26], induction, direct current [27]–[35], and switched reluctance [36]–[43], with specialized design features may be selected depending on the particular needs of the application at hand.

Field oriented control (FOC) of synchronous motor drives involves the transformation of machine quantities from the stationary to the synchronously rotating reference frames in order to convert sinusoidal quantities in the former to constant ones in the latter [44]. This transformation requires an estimate of the instantaneous position of the machine, which is typically achieved using a position sensor. Position sensors, regardless of the specific internal mechanism or the compensation techniques used, exhibit errors, albeit the magnitude of the error varies for different sensing subsystem configurations [45]–[49]. In industrial applications requiring high control performance in terms of torque errors and total torque ripple, it is critical to understand the effects of position estimation errors on the drive system. This is even more important in such applications that are low-cost and involve mass manufacturing applications where the part to part variation may be significant [50]–[55].

Safety requirements often result in both hardware as well as control architecture redundancy for different sensors used in the drive systems. However, manufacturing and cost requirements limit the level of end-of-line compensation that may be employed since the time required for determining compensation parameters starts increasing significantly as the number of sensors increases. Under failure conditions, the drive performance is allowed to be degraded somewhat, but the level of degradation is still required to be contained. When the primary position sensor fails, the drive performance is degraded if the accuracy of the secondary or tertiary sensors is not adequate. During current sensing failures, the drive is operated in feedforward current control mode [56], [57], which is an open-loop control technique, and is more sensitive to position sensing errors as compared to feedback control [58]–[64]. Beyond the torque and current control operation, position and its derivatives including velocity, acceleration and even jerk, are often used for the outer (motion) control loops depending on the application and for implementation of advanced compensation techniques as well as diagnostics. This mandates the need for a comprehensive analysis of the various types of position sensing errors and their effects on the motor drive performance using different control architectures.

Position sensing is one of the most crucial aspects of industrial PMSM drives with high control performance requirements particularly with wide operating conditions in terms of machine speed and torque. For such drives, position sensorless control [65] is not adequate and position sensors such as resolvers or sine-cosine sensors are utilized instead. Even with high precision sensors however, some level of sensing error always exists, sometimes due to inadequate compensation and other times due to limitation of the sensing capabilities inherent in the hardware itself. Position sensing non-idealities or errors may be classified into two primary types, namely static errors or harmonics. Static errors encompass constant sensor offsets or misalignment of the sensor zero with the absolute zero position of the machine and dynamic time delays caused by the sensor magnetics, electrical transmission and analog-to-digital conversion of sensor signals. Harmonics in the position sensing signals arise due to non-linearities in the sensor magnets or misalignments in the relative placement of the rotor shaft, sensor magnet and the sensing circuitry. These non-idealities ultimately result in the degradation of the motor drive performance since the position signal is utilized in several parts of the control system. Further, the effect is different depending on the current control architecture. A comprehensive treatment of position sensing errors, particularly the static errors, describing its effects on motor current and torque tracking, under feedback and feedforward current control operation of PMSM drives has not been presented previously, and is the unique contribution of this paper.

Mathematical (or analytical) models describing the effects of static position sensing errors are presented in this paper. The models are derived in a systematic manner starting from the analysis of sensing and estimation error itself, followed by its effect in the current and torque control. The differences in the performance of feedforward and feedback current control architectures are shown through the derivation of the voltage, current and torque signals. The analysis is presented assuming the current commands are known, i.e., the current reference calculation based on the torque command is not included. This does not lead to any loss of generality in the analysis since the impact of velocity estimation errors on the current command calculation, commonly performed using the well-known maximum torque per ampere (MTPA) and maximum torque per voltage (MTPV) techniques, or more generally determined using advanced power management algorithms [66]–[76], may be included separately as an extension of the presented models.

After the introduction presented in this section, a detailed description of the control architecture of the electric motor drive and its components is provided. The modeling and analysis of position sensing errors along with its impact on the control system exemplified by the expressions of different signals of the drive is presented next. This is followed by elaborate analysis of the performance of motor current control under the presence of position sensing errors. The analysis is verified through experimental results on a 3-phase, 9-slot, 6-pole (9S-6P) permanent magnet synchronous motor (PMSM) drive system rated for 1 kW for a 12 V application.

## Electric Motor Drive Systems

A typical PMSM drive system is shown in Fig. 1. The main hardware components include the motor, the power converter which is a 3-phase voltage source inverter (VSI), phase current and rotor position sensors. The control logic, which illustrates a torque-controlled drive, includes the current command generator, dynamic current controller and inverter commutation blocks. Note that the gate driver which connects the microcontroller to the inverter is not shown explicitly. The torque control of PMSMs is achieved using the well-known FOC technique wherein the stationary reference frame quantities of the electric motor are converted to the synchronous frame using the frame transformation matrix. Torque control achieved this way is achieved indirectly by controlling the motor currents. Mechanistically, this is achieved by converting the measured phase currents to the synchronous frame using the estimated rotor position and then using a current controller developed in the synchronous frame to compute the voltage commands. These voltage commands are converted back to the stationary reference frame by using an appropriate commutation technique along with pulse width modulation. In order to understand the effects of position sensing errors on the PMSM drive performance, the current command generation block is not explained in detail and the analysis is focused to the part of the system between the reference current and generated (actual) torque. Each of these blocks are described and modeled in the following sub-sections.

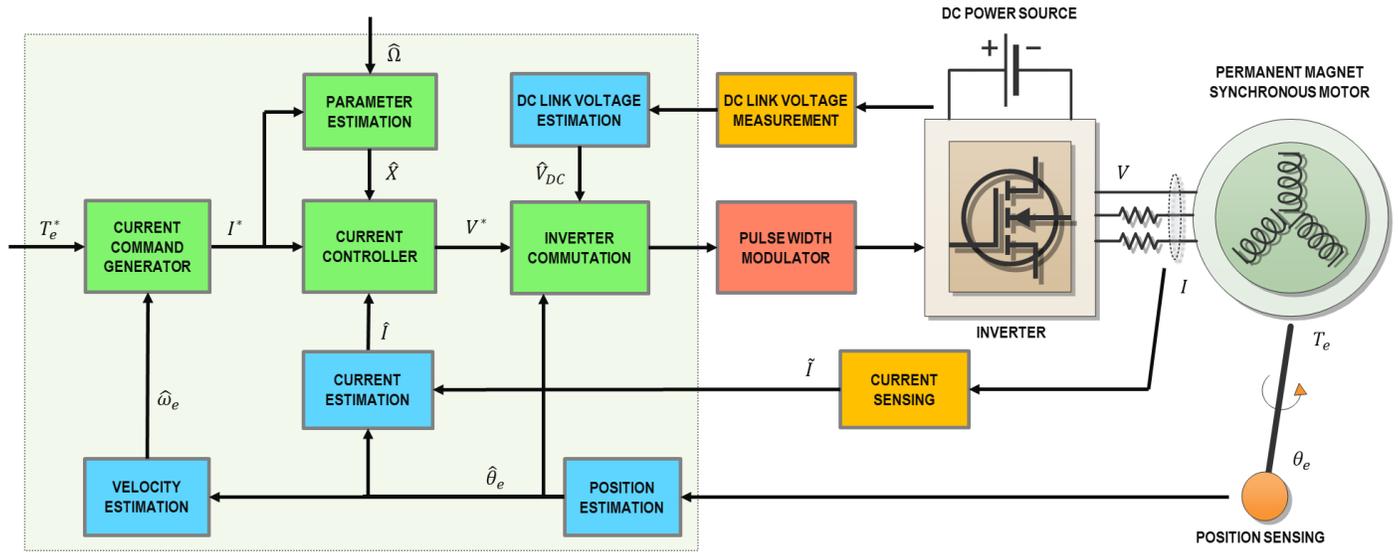

Fig. 1. PMSM drive system block diagram.

## Permanent Magnet Synchronous Machine

The synchronous or dq reference frame model of a PMSM is given in (1) [77], [78].

$$\begin{bmatrix} V_d \\ V_q \end{bmatrix} = \begin{bmatrix} L_d s + R & \omega_e L_q \\ -\omega_e L_d & L_q s + R \end{bmatrix} \begin{bmatrix} I_d \\ I_q \end{bmatrix} + \begin{bmatrix} 0 \\ \omega_e \lambda_m \end{bmatrix}$$

$$T_e = \frac{3}{2} p \big( \lambda_m + (L_q - L_d) I_d \big) I_q$$

(1)

where the subscripts d and q represent the components of the signal vectors in the synchronous reference frame, $s$ donates the derivative operator, $V$ and $I$ are the terminal voltage and current respectively, $L$, $R$ and $\lambda_m$ are the resistance, inductance and PM flux linkage respectively, $\omega_e$ is the rotor electrical velocity, $T_e$ is the electromagnetic torque and $p$ denotes the magnetic pole pairs of the machine. Notice that the zero sequence currents are omitted in (1) since they do not impact the electromagnetic torque generation.

A block diagram representation considering transfer matrices of the machine model in the synchronous reference frame is shown in Fig. 2.

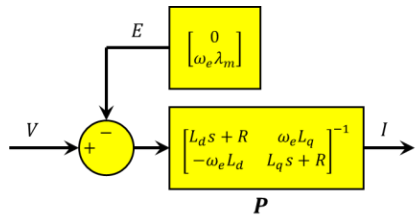

Fig 2. Machine model block diagram.

Two configurations of 9S-6P PMSMs, one having surface mounted magnets (SPMSM) and the other having interior magnets (IPMSM), are considered in this paper. The geometric details and parameters of the SPMSM and IPMSM are given in Table I.

TABLE I. Parameters of 3-phase, 9-slot, 6-pole surface and interior PMSMs.

| Parameter | Value SPMSM | Value IPMSM | Unit |
|---|---|---|---|
| Stack Length | 46.5 | 43 | mm |
| PM Flux Linkage | 7.69 | 7.38 | mWb |
| Inductance d-axis | 59.45 | 102.02 | uH |
| Inductance q-axis | 59.45 | 155.52 | uH |
| Motor Phase Resistance | 6.92 | 9.66 | mΩ |
| FET Resistance | 1.80 | 1.30 | mΩ |
| Winding Type | Concentrated | Concentrated | - |
| Air Gap Length | 7.5 | 5 | mm |
| Rotor Diameter | 36 | 37.1 | mm |
| Stater Outer Diameter | 85 | 85 | mm |
| Tooth Width | 9 | 8.5 | mm |
| Magnet Grade | 42 | 42 | SH |

The cross-sectional views of both these machines are shown in Fig. 3.

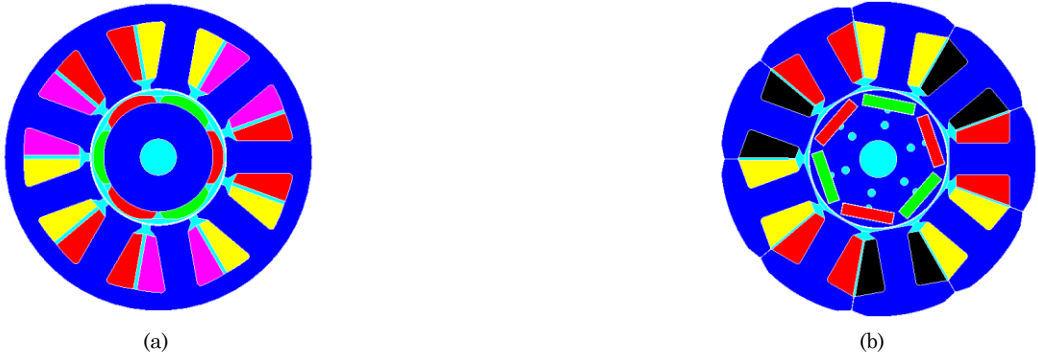

Fig. 3. Cross-sectional view of 3-phase, 9-slot, 6-pole (a) surface (b) interior PMSM with concentrated windings.

The SPMSM is non-salient pole in nature, i.e., the d and q-axis inductances are approximately equal across the operating space. The IPMSM exhibits saliency due to the unequal reluctances in the d and q axis paths of the magnetic circuit. The effect of magnetic saturation is significant in the IPMSM, whereas it is virtually negligible in the case of the SPMSM. In this paper, magnetic saturation is accounted for by scheduling the PM flux linkage and the machine inductances as a function of currents. In particular, the PM flux linkage is assumed to be a function of q-axis current only, while the inductances are assumed to vary with both currents. The saturation maps for the different parameters are illustrated in Fig. 4.

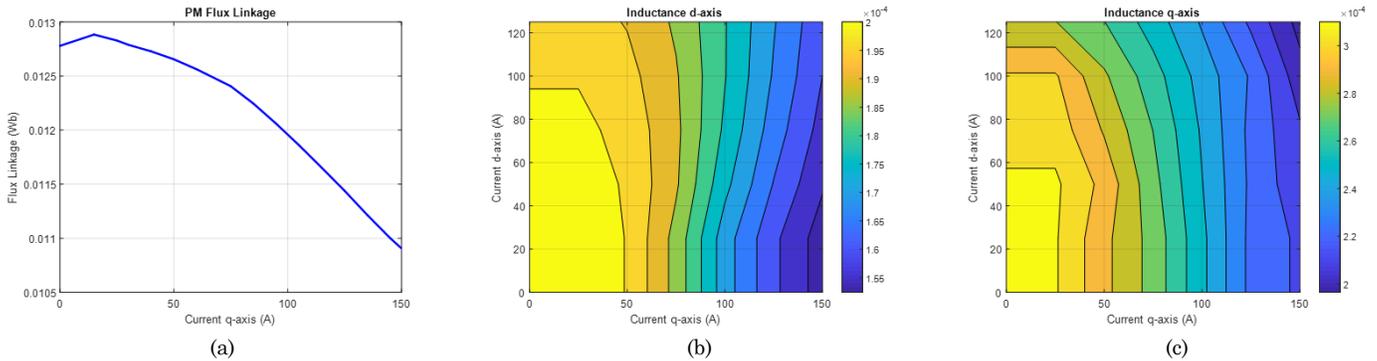

Fig. 4. Saturation maps of (a) PM flux linkage (b) d-axis inductance (c) q-axis inductance for IPMSM.

## Power Converter

The power converter (inverter) is a typical VSI wherein one phase leg is connected to one phase of the motor and the three lower switches in each phase leg are connected to ground. The inverter is operated through the gate drive module which is controlled directly by the microcontroller. The switching action is assumed to be ideal for the presented analysis, i.e., instantaneous switching is assumed, switching non-linearities and transients are ignored, and dead-time effects are neglected. Thus, the actual phase to ground voltages are assumed to be equal to their corresponding commanded values. The switching is performed using the center based or symmetric PWM technique.

Due to the discrete nature of PWM, the commanded inverter voltages are generated in the immediate next PWM period, resulting in a transport lag of one switching period [79]. This phenomenon and the associated mathematical model are described in detail in the next section.

### Current Sensing & Estimation

Current measurement for PMSMs may be performed using a variety of sensor topologies and associated estimation techniques [80]–[86]. In this paper, two in-line phase current measurements are utilized for the estimation of motor currents, one each in phases B and C. A shunt resistor along with a differential operational amplifier (Op-Amp) followed by a first-order hardware RC filter is used for the measurement of phase currents. The measurement is routed to the ADC in the microcontroller.

The total delay in the measurement and transmission of the current signals, including the ADC conversion time, is less than 1 $\mu s$ and is thus neglected. Considering the range of the measurement considering the ADC resolution, the quantization noise is also negligible. Common mode noise caused by the Op-Amps are internally compensated in the microcontroller using offline measurements and are thus not included in the analysis.

### Current Command Generation

This block converts the torque command into d and q-axis current commands by using the well-known MTPA and MTPV algorithms. MTPA involves the calculation of the optimal current commands for a given torque command such that the total motor current command is minimum. While different algorithms are possible for achieving MTPA, a model-based approach utilizing estimated machine parameters is utilized here. In the MTPV region, the calculation of the optimal current commands is performed by considering the voltage constraint, i.e., to maintain the maximum possible torque (as close as possible to the commanded torque) that just meets the voltage command. In general, this is achieved by field weakening where the d-axis current is increased in order to reduce the total induced EMF and the q-axis current is determined, along with the d-axis current, in order to generate the maximum achievable torque.

For non-salient pole machines, the MTPA region results in q-axis current commands only for a given torque command and the d-axis current is only commanded for field weakening under MTPV operation which occurs at higher speeds. In salient pole machine however, d-axis current command is non-zero even at zero speed and increases in the high-speed region when the voltage constraint becomes active. In addition to the basic MTPA and MTPV characteristics, an artificial torque limiter is also imposed before the calculation of the current commands to reduce the torque command below capability. The optimal current commands for both machines are shown in Fig. 5.

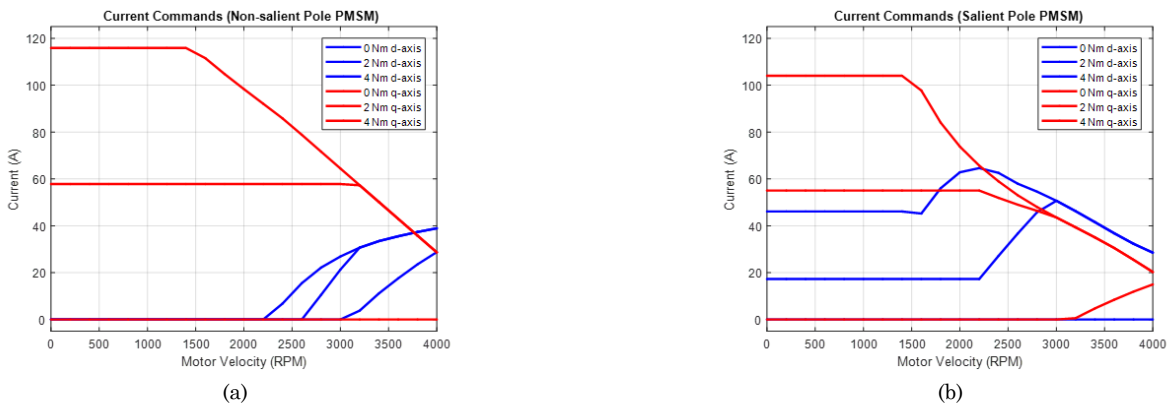

Fig. 5. Current command maps for (a) non-salient pole SPMSM (b) salient pole IPMSM.

In this paper, the analysis is carried out assuming that the current commands are known and thus the current command generation logic is bypassed. This does not lead to any loss of generality in the presented results since the effects of this block can be included separately.

### Parameter Estimation

Due to the highly non-linear nature of PMSMs, the machine parameters vary significantly with temperature and magnetic saturation. The characteristics presented in Fig. 4 are used for incorporating magnetic saturation effects in the parameter estimation block utilizing current commands. Thermal effects are considered for the motor resistance, PM flux linkage and the FET resistances by using temperature estimates of the machine windings, rotor magnets and switches respectively through a lumped thermal model that utilizes the currents to estimate the heat flux and a thermistor placed near the inverter for correction of ambient temperature. The open-loop estimation mechanism used here may be enhanced through the use of online parameter adaptation techniques [87].

### Dynamic Current Control

This block performs the function of generating the synchronous frame voltage commands based on the corresponding current commands. The voltage commands are converted to modulation index and phase advance using the DC link voltage measurement.

As mentioned earlier, current control may be achieved by using either feedforward or feedback control methods. The former uses commanded currents along with an inverse mathematical model of the machine while the latter employs current regulators that utilize measured currents for computing the voltage command values. The detailed description of both these current control methods is presented in a subsequent section.

### Inverter Commutation Technique

A discontinuous space vector PWM (DPWM) technique is used for the commutation of the inverter. The phase duty cycles as a function of electrical position of the motor for maximum modulation are shown in Fig. 6.

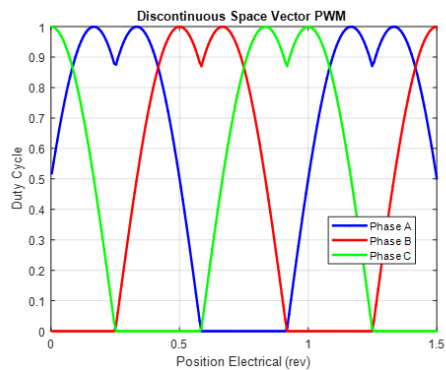

Fig. 6. Duty cycle waveforms for discontinuous space vector PWM.

Although the commutation technique does not affect the proposed analysis pertaining to position sensing errors, it is not described in detail here. However, it is pertinent to mention that a plethora of algorithms may be employed for the control of the power converter [88]–[92]. Further, since DPWM is essentially constructed from the sinusoidal PWM (SPWM) technique but subtracting the minimum duty cycle from each phase for each position, it is sufficient to consider SPWM only for the analysis presented in this paper.

## Position Sensing Model

A sine-cosine type position sensor is used for measuring the rotor position of the PMSM. A detailed description of the construction of such sensors along with an analysis of the non-idealities exhibited by them has been done previously in literature and is thus not repeated here. In general, the raw signals received by the ADC are first corrected for gain, offset and quadrature errors as well as harmonics to obtain the actual sine and cosine signals. These correction terms are usually obtained from offline testing where the sensed signals are compared to a high-resolution position encoder.

The two corrected orthogonal signals are then used in an arctangent function to compute the mechanical position which is then scaled by the number of magnetic pole pairs to obtain the electrical position as expressed in (2).

$$\theta'_e = p \operatorname{atan}^{-1}\left(\frac{u_s}{u_c}\right) \qquad (2)$$

where $u_s$ and $u_c$ represent the corrected sine and cosine signals. Since the placement of the magnet is arbitrary, the zero of the calculated position signal is not aligned to the actual zero of the machine, defined to be the positive zero crossing of the phase A BEMF signal, and thus needs to be adjusted. This is illustrated in Fig. 7 which shows an offset between the fundamental component of phase A BEMF and the sensor position.

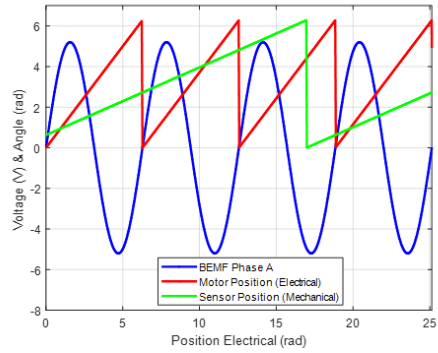

Fig. 7. Phase relationship between sensor position (mechanical), machine position (electrical) and BEMF voltage.

In addition to the sensor offset, due to the delays associated with the magnetics and electrical transmission involved in the position sensing, the measured position signal usually lags the actual position signal. Both these effects may be mathematically expressed as (3).

$$\hat{\theta}'_e(t) = \theta_e(t - t_d) + \delta\theta_0 \qquad (3)$$

where $\delta\theta_0$ is the sensor offset and $t_d$ represents the dynamic time delay involved in position sensing. This relationship may be approximated using Taylor series when the motor speed is constant to obtain (4).

$$\hat{\theta}'_e(t) \approx \theta_e(t) + \omega_e t_d + \delta\theta_0 \qquad (4)$$

Thus, if an estimate of the offset $\delta\hat{\theta}_0$ and the time delay $\hat{t}_d$ is available, the true position of the machine may be obtained using (5).

$$\hat{\theta}_e = \hat{\theta}'_e - \hat{\omega}_e \hat{t}_d - \delta\hat{\theta}_0 \qquad (5)$$

In this paper, the effects of sensor offset and delay is presented in detail, which is typically a consequence of zero or partial compensation of these errors. Thus, the following sections assume no compensation of these errors are performed. However, it should be understood that partial compensation results in similar error signatures in the different signals and therefore the same analysis is applicable to those cases as well. The total position error, consisting of terms due to sensor delay and offset, may be combined as (6).

$$\Delta\theta_e = \omega_e t_d + \delta\theta_0 - \hat{\omega}_e \hat{t}_d - \delta\hat{\theta}_0 \qquad (6)$$

where $\Delta\theta_e$ is the total position error term. Since the current control of the PMSM is performed in the synchronous frame, it is important to understand and model the forward and feedback paths consisting of PWM and current measurement respectively. A timing diagram illustrating the switching of one phase leg and the sampling of current and position signals is shown in Fig. 8.

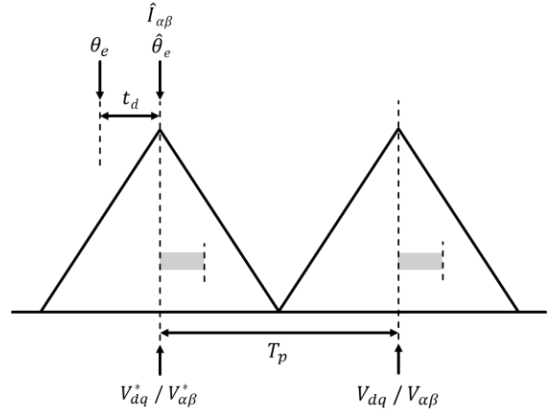

Fig. 8. Timing diagram illustrating PWM and sampling.

It can be seen that the commanded duty cycle is applied in the subsequent PWM cycle resulting in time lag in the voltage application. This behavior may be modeled by computing the instantaneous synchronous frame voltage applied to the machine in terms of the commanded duty cycle as shown in (7).

$$\begin{bmatrix} V_d^* \\ V_q^* \end{bmatrix} = \begin{bmatrix} \cos(\omega_e T_p + \Delta\theta_e) & -\sin(\omega_e T_p + \Delta\theta_e) \\ \sin(\omega_e T_p + \Delta\theta_e) & \cos(\omega_e T_p + \Delta\theta_e) \end{bmatrix} \begin{bmatrix} e^{-T_p s} & 0 \\ 0 & e^{-T_p s} \end{bmatrix} \begin{bmatrix} V_d \\ V_q \end{bmatrix} \tag{7}$$

The measured phase currents in the stationary frame are used with the sampled position to estimate the synchronous frame currents as (8).

$$\begin{bmatrix} \hat{I}_d \\ \hat{I}_q \end{bmatrix} = \begin{bmatrix} \cos\Delta\theta_e & -\sin\Delta\theta_e \\ \sin\Delta\theta_e & \cos\Delta\theta_e \end{bmatrix} \begin{bmatrix} I_d \\ I_q \end{bmatrix} \tag{8}$$

## Current Control Techniques

Dynamic current control of PMSMs is performed using either feedforward or feedback techniques as mentioned earlier and are described in detail in this section.

### Feedforward Current Control

Feedforward current control involves computing the voltage commands using an inverse model of the motor with estimated machine parameters as (9).

$$\begin{bmatrix} V_d^* \\ V_q^* \end{bmatrix} = \begin{bmatrix} \hat{L}_d \hat{s} + \hat{R} & \omega_e \hat{L}_q \\ -\omega_e \hat{L}_d & \hat{L}_q \hat{s} + \hat{R} \end{bmatrix} \begin{bmatrix} I_d^* \\ I_q^* \end{bmatrix} + \begin{bmatrix} 0 \\ \hat{\omega}_e \hat{\lambda}_m \end{bmatrix} \tag{9}$$

where $\hat{s}$ represents an estimate of the derivative operator. A block diagram representation of the feedforward current controller is shown in Fig. 9.

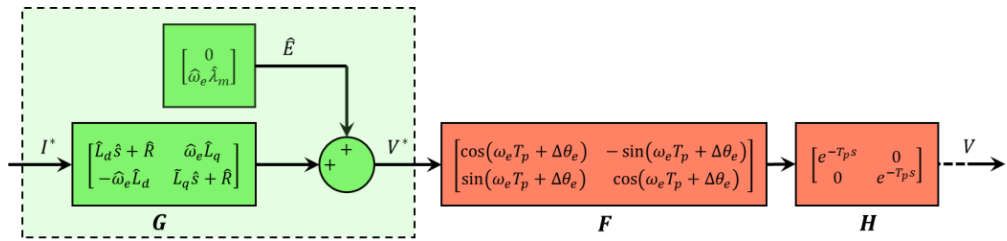

Fig. 9. Feedforward current control block diagram.

While several methods for approximating the derivative operation are possible, a pure derivative along with a low pass filter with a time constant $\tau_f$ is considered in this case as given in (10).

$$\hat{s} = \frac{s}{\tau_f s + 1} \tag{10}$$

The dynamic feedforward current controller is described thus far. If the dynamic terms are set to zero, the resulting compensator is called a static feedforward controller since it does not account for the dynamics of the plant model and is expressed as (11).

$$\begin{bmatrix} V_d^* \\ V_q^* \end{bmatrix} = \begin{bmatrix} \hat{R} & \omega_e \hat{L}_q \\ -\omega_e \hat{L}_d & \hat{R} \end{bmatrix} \begin{bmatrix} I_d^* \\ I_q^* \end{bmatrix} + \begin{bmatrix} 0 \\ \hat{\omega}_e \hat{\lambda}_m \end{bmatrix} \tag{11}$$

Due to the high dependency of the feedforward compensator on the estimation of plant parameters, it usually tends to be less robust to modeling uncertainties (i.e., estimation errors). Further, this control method exhibits relatively low bandwidth and has poor disturbance rejection capability. In the case of PMSM current control, disturbance rejection primarily refers to the compensation of the BEMF voltage term.

The inaccurate estimation of position results in non-zero current tracking error which in-turn causes erroneous torque generation. The mathematical expressions capturing this behavior are obtained as (12).

$$I = (PHFG)I^* + P(HF\hat{E} - E) \tag{12}$$

This expression may be simplified by assuming that the machine parameter estimation is accurate to yield the steady state current outputs as (13).

$$\begin{bmatrix} I_d \\ I_q \end{bmatrix} = \begin{bmatrix} \cos \Delta \theta_e' - \frac{\omega_e R(L_q - L_d)}{R^2 + \omega_e^2 L_d L_q} \sin \Delta \theta_e' & -\sin \Delta \theta_e' \\ \sin \Delta \theta_e' & \cos \Delta \theta_e' + \frac{\omega_e R(L_q - L_d)}{R^2 + \omega_e^2 L_d L_q} \sin \Delta \theta_e' \end{bmatrix} \begin{bmatrix} I_d^* \\ I_q^* \end{bmatrix} \\ + \frac{1}{R^2 + \omega_e^2 L_d L_q} \begin{bmatrix} -R \sin \Delta \theta_e' + \omega_e L_q (1 - \cos \Delta \theta_e') \\ -R(1 - \cos \Delta \theta_e') - \omega_e L_d \sin \Delta \theta_e' \end{bmatrix} \omega_e \lambda_m \tag{13}$$

where $\Delta \theta_e' = \omega_e T_p + \Delta \theta_e$. A simplified expression for non-salient pole machines with a synchronous inductance of $L_0$ is given in (14).

$$T_e = \frac{3}{2} p \lambda_m \left( I_q^* \cos \Delta \theta_e' + I_d^* \sin \Delta \theta_e' - \frac{\omega_e^2 L_0 \lambda_m}{R^2 + \omega_e^2 L_0^2} \sin \Delta \theta_e' + \frac{\omega_e^2 L_0 R}{R^2 + \omega_e^2 L_0^2} (1 - \cos \Delta \theta_e') \right) \tag{14}$$

The torque command may be expressed in terms of the current commands as (15).

$$T_e^* = \frac{3}{2} p (\hat{\lambda}_m + (\hat{L}_q - \hat{L}_d) I_d^*) I_q^* \tag{15}$$

### Feedback Current Regulation

Feedback current regulators employ proportional-integral (PI) controllers on the error between the d and q axis current errors along with a feedforward BEMF compensation term as shown in Fig. 10. In addition to this, a decoupling network may additionally be employed to reduce the cross-coupling effects between the two current loops. Decoupling networks are not considered in the subsequent description, but the presented analysis may easily be extended to include their effects.

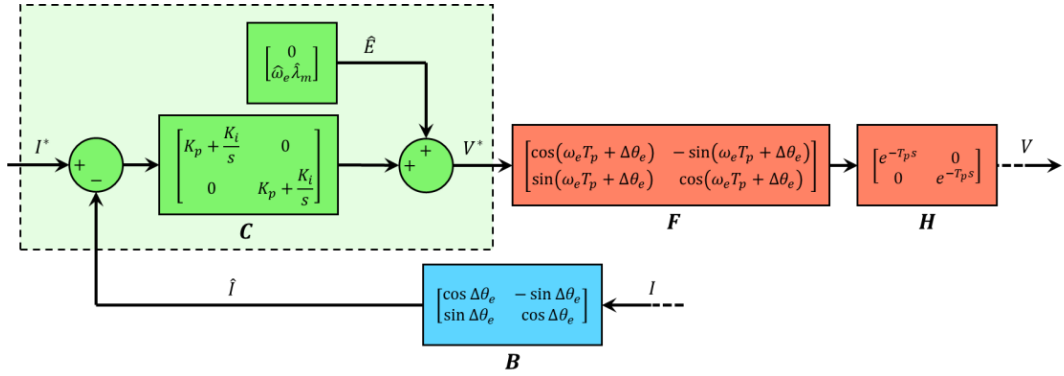

Fig. 10. Feedback current regulation block diagram.

In the case of feedback regulators, the transfer matrices representing the closed loop current control behavior are obtained as (16).

$$I = (J + PHFCB)^{-1}I^* + (J + PHFCB)^{-1}P(HF\hat{E} - E) \quad (16)$$

where $J$ is the identity matrix. In steady state, this expression gets simplified to (17).

$$\begin{bmatrix} I_d \\ I_q \end{bmatrix} = \begin{bmatrix} \cos \Delta\theta_e & \sin \Delta\theta_e \\ -\sin \Delta\theta_e & \cos \Delta\theta_e \end{bmatrix} \begin{bmatrix} I_d^* \\ I_q^* \end{bmatrix} \quad (17)$$

The electromagnetic torque is thus obtained as (18).

$$T_e = \frac{3}{2}p\lambda_m \left[ (I_q^* \cos \Delta\theta_e + I_d^* \sin \Delta\theta_e) + (L_q - L_d)\left( I_d^* I_q^* \cos(2\Delta\theta_e) + \frac{I_d^{*2} - I_q^{*2}}{2}\sin(2\Delta\theta_e) \right) \right] \quad (18)$$

Since the current regulators inherently compensate (partially) for the position sensing induced errors in the forward path, the information related to these errors shows up in the commanded voltage signals. When the current commands are set to zero, the voltage commands are given by (19).

$$\begin{bmatrix} V_d^* \\ V_q^* \end{bmatrix} = \begin{bmatrix} \cos \Delta\theta_e' & \sin \Delta\theta_e' \\ -\sin \Delta\theta_e' & \cos \Delta\theta_e' \end{bmatrix} \begin{bmatrix} 0 \\ \omega_e \lambda_m \end{bmatrix} \quad (19)$$

This equation shows that when the commanded currents are zero, the voltage signals are directly related to the machine parameters, motor velocity and the position sensing errors. Thus, by commanding zero current commands and by changing velocity in steps, the total error term at each velocity may be calculated from which the individual errors, namely the offset and delay error terms, can be extracted. An alternative method to estimate the error terms is by using the torque measurement when only d-axis current is commanded since that would result in generation of torque even though the commanded torque is zero (due to the absence of q-axis current command).

Before concluding this section, it is pertinent to mention that while the technique for obtaining the error (and thus compensation) terms is described in the context of feedback current control, the same idea may be applied in feedforward control to obtain these errors as well. In that case, the difference between estimated currents in the synchronous frame and the ideal currents, obtained from the knowledge of the plant parameters and commanded voltages, may be used to compute the errors.

## Validation

The experimental setup used for validating the analysis presented thus far is described in this section. An induction motor (IM) drive is used as the servomotor to rotate the test motors at constant speed and a torque sensor and a position encoder are used to obtain the data using a personal computer. The test motor is supplied from a 12 V power supply with a current sourcing capability of 200 A. The PMSM drive under test is controlled from a separate computer, which is used to directly command synchronous frame currents.

The SPMSM and IPMSM drives are tested in feedback and feedforward current control modes at different torque and speed points under both uncompensated and compensation position sensor error conditions. The test results are compared with the analytical results and presented in the foregoing description. For the sake of brevity, only the measured torque values for all the test conditions are presented. Note that the PWM frequency is 16 kHz, equivalent to a switching period of 62.5 $\mu s$ and the sensor delay is 52.5 $\mu s$ and 40 $\mu s$ for the SPMSM and IPMSM respectively.

## Feedforward Current Control

### Sensor Delay

The experimental results for the non-salient pole SPMSM operated in feedforward current control mode with sensing delays with different levels of compensation at torque commands of 0 Nm and 4 Nm are shown in Fig. 11. The current commands are zero with a torque command of 0 Nm and thus the BEMF term is dominant resulting in a linear decrease of the q-axis current and torque in the low to medium region. When the speed increases and reaches 3200 RPM, the machine enters field weakening operation and d-axis current command becomes zero. This causes the q-axis current and torque to become virtually constant. At 4 Nm, the effect on q-axis current and torque is similar to that at 0 Nm, but since the q-axis current command is non-zero, the actual d-axis current decreases till the field weakening region is reached at 2200 RPM after which the d-axis current increases and the error in q-axis current and torque levels out. The errors reduce as the level of compensation is increased. The salient pole IPMSM exemplifies very similar behavior as shown in Fig. 12 but shows a continues decrease in torque in the field weakening region. This is because as d-axis current is generated, the reluctance torque becomes non-zero under field weakening operation.

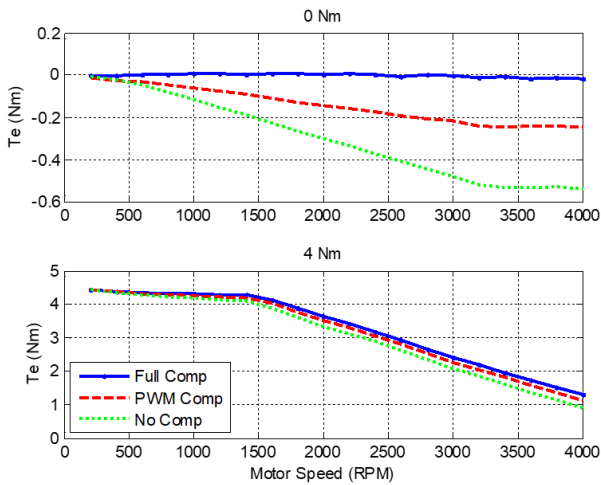

Fig. 11. Torque output of SPMSM under feedforward current control with position sensor delay at different compensation levels with 0 Nm (top) and 4.5 Nm (bottom) commands.

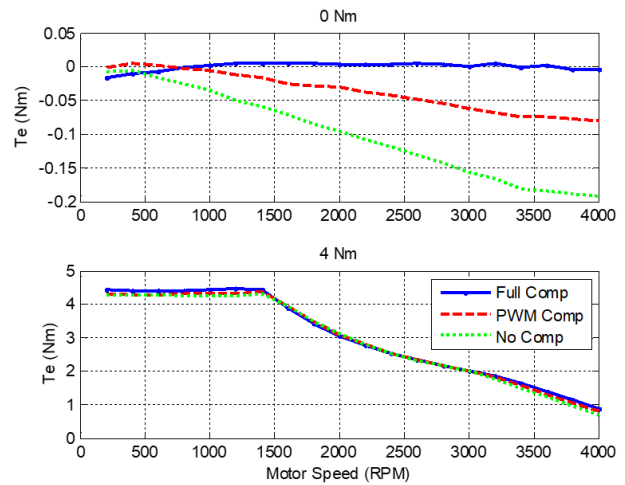

Fig. 12. Torque output of IPMSM under feedforward current control with position sensor delay at different compensation levels with 0 Nm (top) and 4.5 Nm (bottom) commands.

### Sensor Offset

The steady state performance of the SPMSM drive with positive and negative sensor offset of $15^0$ electrical are shown in Fig. 13. The sinusoidal signature of the error is very evident in the results because the torque follows q-axis current behavior, while the d-axis current, which actually is asymmetric for positive and negative errors (not shown) does not impact the torque output. In particular, the d-axis current for a negative offset first decreases and becomes negative in the low speed region and then gradually increases again with increasing speed and becomes virtually equal to the commanded value at very high speeds. This behavior, slightly exaggerated in that the current actually increases beyond the commanded value in the case of IPMSMs as shown in Fig. 14, causes the torque output to be flatter due to the reluctance torque generation.

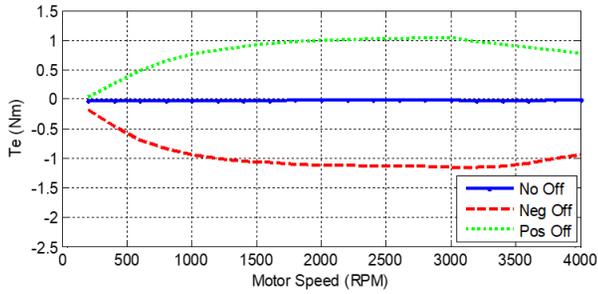

Fig. 13. Torque output of SPMSM under feedforward current control with positive and negative position sensor offset (misalignment) with 0 Nm command.

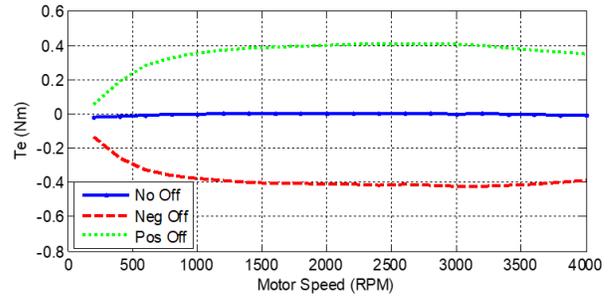

Fig. 14. Torque output of IPMSM under feedforward current control with positive and negative position sensor offset (misalignment) with 0 Nm command.

## Feedback Current Regulation

### Sensor Delay

The effect of sensor delay in SPMSMs under feedback current control, as shown in Fig. 15, is a non-zero torque output with 0 Nm command in the high-speed range. This can be attributed to the fact that the machine enters field weakening at higher speeds resulting in a non-zero d-axis current command which in-turn causes an errors q-axis current command and thus torque to be generated. At a command of 4 Nm, the torque error is consistent throughout the operation range and the actual torque exceeds the command. The IPMSM drive shows similar behavior at 0 Nm as illustrated in Fig. 16. However, at a torque command of 4 Nm, the torque output is below the command in the MTPA range and increases in the field weakening region to exceed the commanded value. Note that the slight deviation from zero of the torque plots is due to measurement non-linearities and imperfect friction correction.

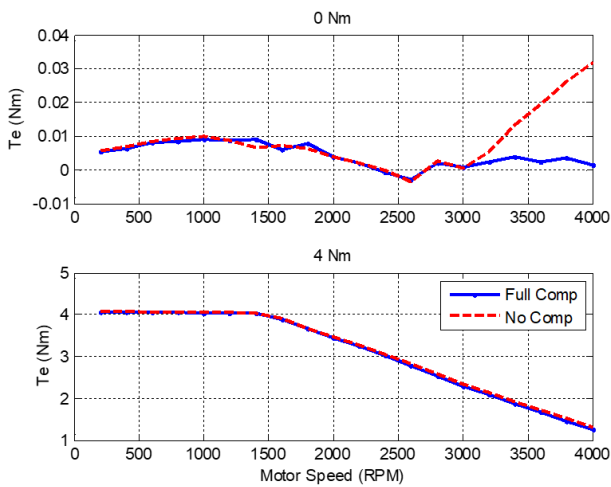

Fig. 15. Torque output of SPMSM under feedback current control with position sensor delay with and without compensation with 0 Nm (top) and 4 Nm (bottom) commands.

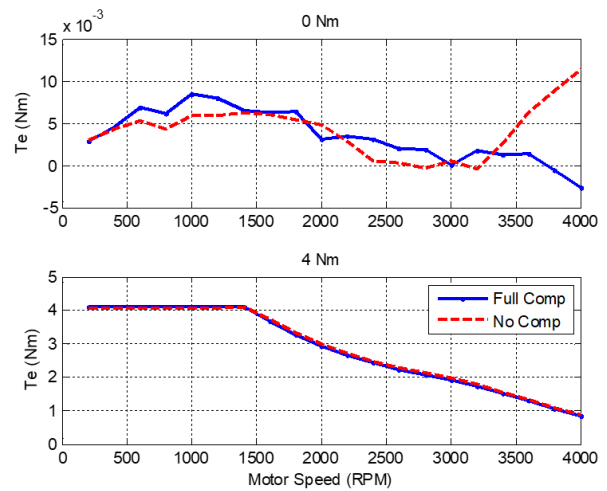

Fig. 16. Torque output of IPMSM under feedback current control with position sensor delay with and without compensation with 0 Nm (top) and 4 Nm (bottom) commands.

### Sensor Offset

The effects of sensor offset errors, as shown in Fig. 17, on SPMSMs is very similar to that of delays in feedback current control at 0 Nm. This is attributed to the fact that the error is relatively low and so the sinusoidal terms change approximately linearly at the speeds at which the errors show up and due to the fact that the range of speeds tested is limited. The torque output shows a slight non-linearity at around 3700 RPM which is due to a torque sensor measurement issue at that specific point and may be disregarded. The behavior of IPMSM drives, illustrated in Fig. 18, is also very similar to SPMSMs, but this is only the case at 0 Nm torque command with sensor offset errors.

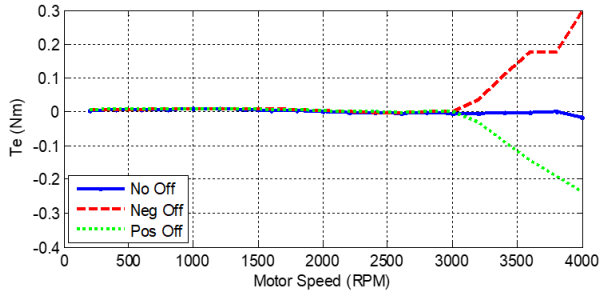 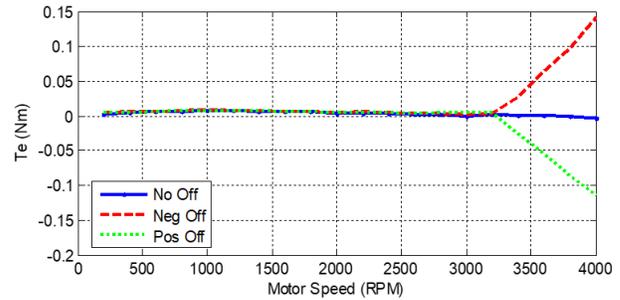

Fig. 17. Torque output of SPMSM under feedback current control with positive and negative position sensor offset (misalignment) with 0 Nm command.

Fig. 18. Torque output of IPMSM under feedback current control with positive and negative position sensor offset (misalignment) with 0 Nm command.

## Conclusion

The impact of non-idealities in position sensing, particularly time delays and offsets, on the performance of PMSM drives is presented in this paper. The mathematical framework for modeling position sensing errors is developed and then used to understand and explain the effects of these errors on both feedforward and feedback current control architectures. The analytical results are extensively validated with experimental tests on both non-salient and salient pole PMSM drive systems.